\documentclass[useAMS,usenatbib]{mn2e}

\usepackage{graphicx}
\usepackage{txfonts}
\usepackage{natbib}

\title[Probing the nature of IGR J16493$-$4348]{Probing the nature of IGR J16493$-$4348: \\Spectral and temporal analysis of the 1--100 keV emission}

\author[A. B. Hill]{A. B. Hill$^{1}$\thanks{E-mail:abh@astro.soton.ac.uk} 
, A. J. Dean$^{1}$, R. Landi$^{2}$, V. A. McBride$^{1}$, A. De Rosa$^{3}$,  
\newauthor A. J. Bird$^{1}$, A. Bazzano$^{3}$,V. Sguera$^{2}$ \\
$^{1}$School of Physics \& Astronomy, University of Southampton, UK\\
$^{2}$IASF-INAF, via P. Gobetti 101, Bologna 40129, Italy\\
$^{3}$IASF-INAF, via del Fosso del Cavaliere 100, Roma 00113, Italy\\}
\begin{document}

\date{Accepted/Received}

\pagerange{\pageref{firstpage}--\pageref{lastpage}} \pubyear{2007}

\maketitle
\label{firstpage}

\begin{abstract}
IGR J16493$-$4348 was one of the first new sources to be detected by the INTEGRAL $\gamma$-ray telescope in the 18--100 keV energy band.  Based upon spatial coincidence the source was originally associated with the free radio pulsar PSR J1649$-$4349.  Presented here are the results of 2.8 Ms of observations made by the INTEGRAL mission and a 5.6 ks observation with the Swift X-ray Telescope.  Spectral analysis indicates that the source is best modeled by an absorbed power law with a high energy cut-off at E$_{cut}$$\sim$15 keV and a hydrogen absorbing column of N$_H$=5.4$^{+1.3}_{-1}$~$\times$~10$^{22}$~cm$^{-2}$. Analysis of the light curves indicates that the source is a weak, persistent $\gamma$-ray emitter showing indications of variability in the 2--9 and 22--100 keV bands.  The average source flux is $\sim$1.1~$\times$~$10^{-10}$erg~cm$^{-2}$~s$^{-1}$ in the 1--100 keV energy band.  No coherent timing signal is identified at any timescale in the INTEGRAL or Swift data.

The refined source location and positional uncertainty of IGR J16493$-$4348 places PSR J1649$-$4349 outside of the 90\% error circle.  We conclude that IGR J16493$-$4348 is not associated with PSR J1649$-$4349.  Combining the INTEGRAL observations with Swift/XRT data and information gathered by RXTE and Chandra we suggest that IGR J16493$-$4348 is an X-ray binary; and that the source characteristics favour a high mass X-ray binary although an LMXB nature cannot be ruled out.
\end{abstract}


\section{Introduction}
Since its launch in 2002 the INTErnational Gamma-Ray Astrophysics Laboratory (INTEGRAL) has been making regular scans of the Galactic plane and the Galactic centre as part of its Core Programme \citep{2003A&A...411L...1W}.  The Imager on-Board INTEGRAL (IBIS) generates images of the sky with a $\sim$30$^{\circ}$ field of view, an angular resolution of 12$\arcmin$ FWHM and a point source location accuracy of 30$\arcsec$--3$\arcmin$ in the 15-1000 keV energy range \citep{2003A&A...411L.131U}.  Using the soft $\gamma$-ray ISGRI detector of IBIS, a number of surveys of the Galactic plane have been published spanning the 18--100 keV energy range \citep{Bird04-cat1, Bird06-cat2, Bird07-cat3}. The 1$^{st}$ IBIS/ISGRI survey, \citet{Bird04-cat1}, announced the detection of PSR J1649$-$4349, a 0.87s radio pulsar, at 6.3~$\sigma$ in the 30--50 keV energy range.  \citet{ATel457}, also detected the source with INTEGRAL during a deep exposure of the Norma spiral arm tangent at 9.1~$\sigma$ in the 18--45 keV band at R.A.~=~16$^{h}$~49$^{m}$~21$^{s}$ Dec.~=~$-$43$^{\circ}$~48$\arcmin$~36$\arcsec$ with a 2$\arcmin$ positional uncertainty.  They suggested that the source may be an unknown X-ray binary and that the pulsar is serendipitous to the detected source.  In the subsequent IBIS/ISGRI catalogues released by Bird et al. the source is listed as IGR J16493$-$4348 due to the uncertainty over the nature of the source.

RXTE observations were made of IGR J16493$-$4348 using the PCA instrument by \citet{ATel465} in April 2005.  Two observations were performed on the 14 and 15 April for 3.5 and 2.6 ks respectively.  No significant pulsations were detected at any timescale including the known 0.87s period of PSR J1649$-$4349.  \citet{ATel465} find that the mean spectrum of IGR J16493$-$4348 is consistent with a heavily absorbed power law with N$_H$~$\sim$~10$^{23}$ cm$^{-2}$ and a photon index of 1.4.  They find a flux of 1.0$\times$10$^{-11}$, 1.3$\times$10$^{-11}$ and 2.1$\times$10$^{-11}$ erg cm$^{-2}$ s$^{-1}$ in the 2--10, 10--20 and 20--40 keV energy ranges.

Chandra imaging of the field of IGR J16493$-$4348 in October 2005 was performed for 4.1 ks by \citet{ATel654}.  They found that within the 2$\arcmin$ error circle of the INTEGRAL source a single point source was detected at R.A.~=~16$^{h}$~49$^{m}$~26.92$^{s}$ Dec.~=~$-$43$^{\circ}$~49$\arcmin$~8.96$\arcsec$, with a 0.6$\arcsec$ error in each coordinate.  This position is 1.2$\arcmin$ away from the position of PSR J1649$-$4349. \citet{ATel654} associate this source with IGR J16493$-$4348 and no X-ray photons are detected within 5$\arcsec$ of the pulsar position.  The Chandra light curve showed indications of variability but no periodicities were found in the 1--500s range and no X-ray spectrum could be extracted from the data.  \citet{ATel654} note that the earlier spectral measurement of IGR J16493$-$4348 by RXTE may be contaminated by another X-ray source, 1RXS J164913.6-435527, as this source is $\sim$6.7$\arcmin$ away and would be within the field of view of the PCA and was observed to be a bright 22.4$\sigma$ source in the Chandra image.

IGR J16493$-$4348 is located on the Galactic plane, 20$^{\circ}$ from the Galactic centre in the vicinity of the Norma spiral arm tangent. This location and the high absorbing hydrogen column density measured by RXTE make this INTEGRAL source a candidate of the class of highly obscured high mass X-ray binaries (HMXBs) of which INTEGRAL has discovered numerous examples \citep{2005AIPC..797..402K,2005A&A...443..485D, 2006A&A...453..133W, 2007A&A...467..585B}.

In this paper we present a complete analysis of all archival INTEGRAL observations of IGR J16493$-$4348 and an analysis of Swift X-ray Telescope observations of the source in March 2006.  Spectral and temporal analysis of both data sets is reported.

\section{Observations}

\subsection{INTEGRAL data}
The entire INTEGRAL data archive of public and Core Programme was searched for pointings which included IGR J16493$-$4348.  A range of observations were found from 28 February 2003 -- 30 July 2006.  3843 individual pointings were made with IGR J16493$-$4348 in the field of view, which corresponds to an effective exposure of 2.8 Ms.  The individual IBIS/ISGRI pointing data were reduced using the {\sc offline standard analysis}, {\sc osa} software version 5.1.  A light curve of the individual pointings, each with an exposure of $\sim$2 ks, was produced and an image mosaic was constructed using the proprietary software of \citet{Bird07-cat3}.

The maximum source detection in a single IBIS pointing is $\sim$4.2$\sigma$ and typically it is considerably less than this; consequently it is not possible to extract a good spectrum from any individual pointing.  All 3843 pointings were processed with the {\sc osa} software to produce images in 12 energy channels spanning the 22--300 keV range; the channels in the 22--100 keV range were logarithmically spaced to evenly distribute the counts across the channels (assuming a Crab-like spectrum). The weighted average flux of the source in each energy channel was calculated from the individual {\sc osa} flux and variance images and used to construct a standard spectral \textit{pha} file.  An appropriate rebinned \textit{rmf} file was produced from the standard IBIS spectral response file to match the 12 chosen energy channels.  This production of the spectra of weak persistent sources with INTEGRAL data by imaging in fine energy bands is an established method \citep{2007MNRAS.tmp..709L, 2006MNRAS.371..821M, 2005A&A...439..255H} and has been extensively tested against the 'standard' analysis for bright sources.  This technique is fundamentally the same as that described in Appendix A of the spectral analysis of IGR J19140+0951 by \citet{2005A&A...432..235R} and is the same technique applied in the {\sc osa} \textit{mosaic\_spec} tool.

\subsection{SWIFT/XRT data}

IGR J16493$-$4348 was observed with XRT (X-ray Telescope, 0.2--10 keV) on board the Swift satellite (Gehrels et al. 2004) between 23:11:34 UTC 11 March 2006 and 09:03:54 UTC 12 March 2006 for a total exposure of 5.6 ks. Data reduction was performed using the XRTDAS v.2.0.1 standard data pipeline package ({\sc xrtpipeline} v. 0.10.6), in order to produce screened event files. All data were extracted only in the Photon Counting (PC) mode \citep{2004SPIE.5165..217H}, adopting the standard grade filtering (0--12 for PC) according to the XRT nomenclature. Events for spectral and temporal analysis were extracted within a circular region of radius 20$^{\prime \prime}$, centered on the source position, which encloses approximately 90\% of the PSF at 1.5 keV \citep{2004SPIE.5165..232M}. The source background was measured within a source-free circular region with radius 90$^{\prime \prime}$ located far from the source. Spectra and light curves were extracted from the corresponding event files using the {\sc xselect} v.2.4 software. The spectra were binned using {\sc grppha} in an appropriate way, so that the $\chi^{2}$ statistic could be applied. We used the latest version (v.009) of the response matrices and created individual ancillary response files (ARF) using {\sc xrtmkarf} v.0.5.2. 

\section{Analysis}

\begin{figure}
	\centering
	\includegraphics[width=0.98\linewidth, clip]{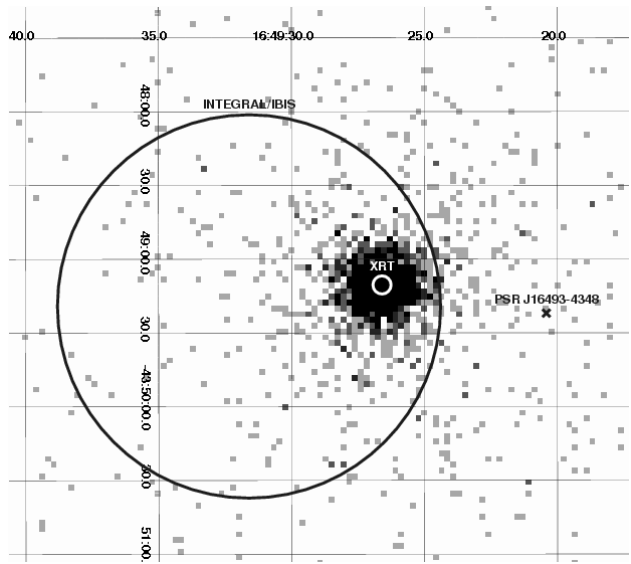}
	\caption{The Swift/XRT X-ray image of the field around PSR J1649$-$4349.  The cross indicates the position of PSR J1649$-$4349.  The circle on the left is the position of the INTEGRAL source found in this paper with a 1.3$\arcmin$, 90\% error circle.  The XRT 90\% error circle is only 3.6$\arcsec$ in radius and is located inside the INTEGRAL error circle.  The Chandra position \citep{ATel654} is within the XRT error circle. The image coordinates are in R.A. and Dec. and the field is approximately 5$\arcmin \times$ 4$\arcmin$.}
	\label{fig:image}
\end{figure}

Upon inspecting the 18--60 keV mosaic of all of the INTEGRAL/IBIS pointings, a 21.4$\sigma$ detection of the source is made with a corresponding flux of 2.3~$\pm$~0.1 mCrab.  The extracted source position is R.A.~=~16$^{h}$ 49$^{m}$~31.6$^{s}$ Dec.~=~$-$43$^{\circ}$~49$\arcmin$~19.2$\arcsec$, with a 1.3$\arcmin$ 90\% error circle.  The error circle is calculated using the formulation of \citet{2003A&A...411L.179G} which has been shown to be representative of INTEGRAL source detections by \citet{Bird07-cat3}.  The location of PSR J1649$-$4349 is R.A.~=~16$^{h}$~49$^{m}$~20.42$^{s}$ Dec.~=~$-$43$^{\circ}$~49$\arcmin$~22$\arcsec$ \citep{2001MNRAS.328...17M}.  The position of the INTEGRAL detection of IGR J16493$-$4348 and the position of PSR J1649$-$4349  are shown superimposed upon the Swift image in Figure~\ref{fig:image}.  The increased amount of INTEGRAL data and the correspondingly higher significance of detection has resulted in a more precise source position of IGR J16493$-$4349.  Whilst our position is compatible with that found for the original detection \citep{ATel457} and the Chandra detection \citep{ATel654}, the location of PSR J1649$-$4349 is now outside of the 90\% error circle and is $\sim$2$\arcmin$ from the refined position of IGR J16493$-$4348.

The Swift/XRT image of the field yields a source detection of 45.9$\sigma$ in the 1--9 keV energy band.  The XRT source position is derived with the {\sc xrtcentroid} task v.0.2.7 and including the correction for the misalignment between the telescope and the satellite optical axis \citep{2006A&A...448L...9M} is R.A.~=~16$^{h}$ 49$^{m}$~26.6$^{s}$ Dec.~=~$-$43$^{\circ}$~49$\arcmin$~10.5$\arcsec$, with a 3.6$\arcsec$ 90\% error circle.  The Swift position is shown in Figure~\ref{fig:image} and is compatible with both the new IBIS position reported here and the reported Chandra position.  Within a 5$\arcsec$ circle centred on the known position of pulsar PSR J1649$-$4349 the Swift/XRT detects a single X-ray photon within the 5.6 ks exposure.  This is compatible with the non-detection of PSR J1649$-$4349 by Chandra which detected no photons in an area of the same size \citep{ATel654}. 

\subsection{Timing Analysis}
An inspection of the light curve of the individual IBIS pointings, each with an exposure of $\sim$2 ks, indicates that IGR J16493$-$4348 is a weak, persistent source with no individual pointing having a detection significance of 5$\sigma$.  A timing analysis was performed using the Lomb-Scargle periodogram method \citep{1982ApJ...263..835S, 1989ApJ...338..277P}; no coherent signal was detected in the 2 ks -- 400 day period range.  The weak nature of the source would require a very high amplitude periodic signal to be present in order to be clearly detected, however a large amplitude signal would be expected if the system were undergoing regular eclipses of a donor star.  Additionally the weak detection of the source on 2~ks times precluded the possibility of searching on shorter timescales for any indication of pulsations.

Investigating the general trend of the long term IBIS light curve provides some evidence of long term variability in the source.  The weighted mean flux of the source was fitted to the light curve and yielded a reduced chi-square value of $\chi^2_\nu$ = 1.3 for 3841 degrees of freedom.  Whilst not an unacceptable fit, the deviation of the chi-square value from 1.0 indicates some level of variability is present in the light curve although it is not of a significant amplitude to be characterised further.

Light curves with 5s and 50s binning were produced in the 1--9 keV band from the Swift/XRT data.  The Lomb-Scargle periodogram method was again used to seach for any periodic signals present in the data.  No coherent timing signal from 10--1000s was evident in the analysis.  Using a Monte-Carlo approach, it was possible to assess the data quality of the 5s binned light curve and our sensitivity to periodic signals. Simulated light curves were generated which had the same sampling and statistical properties of the data but which had a sinusoidal modulation of $\sim$100 s introduced. The results of the Monte-Carlo simulations indicated that if the source has a pulse amplitude of $<$~35\% we would be unable to detect it.  Whilst no coherent signal was evident the 50s binned XRT light curve was not well fit by simple continuous emission.  A fit to the weighted mean of the light curve yielded a reduced chi-square value of $\chi^2_\nu$ = 1.7 indicating that the source flux was variable during the observation.

\subsection{Spectral Analysis}

\begin{figure}
\centering
\includegraphics[width=0.95\linewidth, clip]{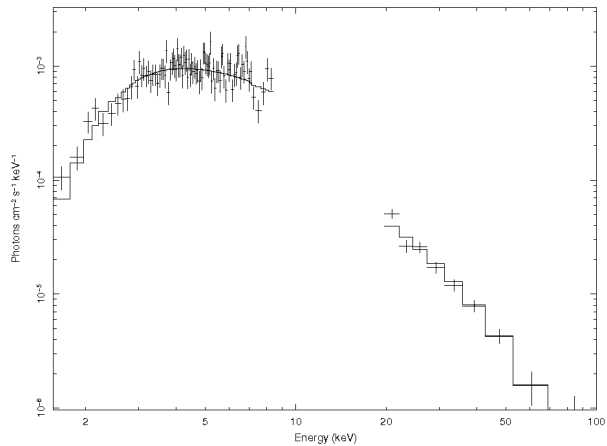}
\caption{The unfolded energy spectrum of the combined Swift/XRT and INTEGRAL/ISGRI spectra of IGR J16493$-$4348.  The spectra have been fit with an absorbed high energy cut-off power law model. (see Table~\ref{tab:spec}).}
\label{fig:spec}
\end{figure}

\begin{figure*}
\centering
\includegraphics[width=0.48\linewidth, clip]{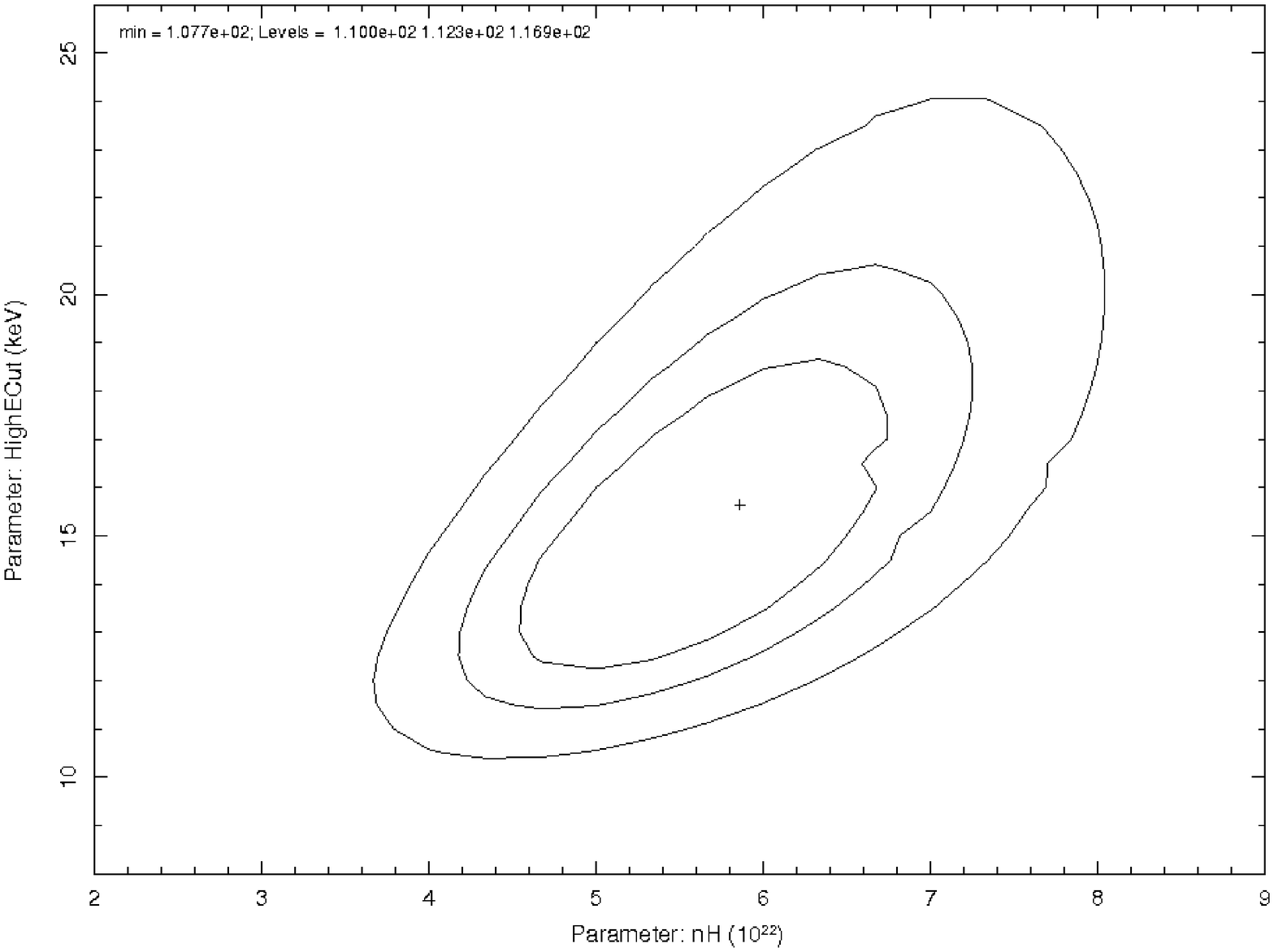}
\hfill
\includegraphics[width=0.48\linewidth, clip]{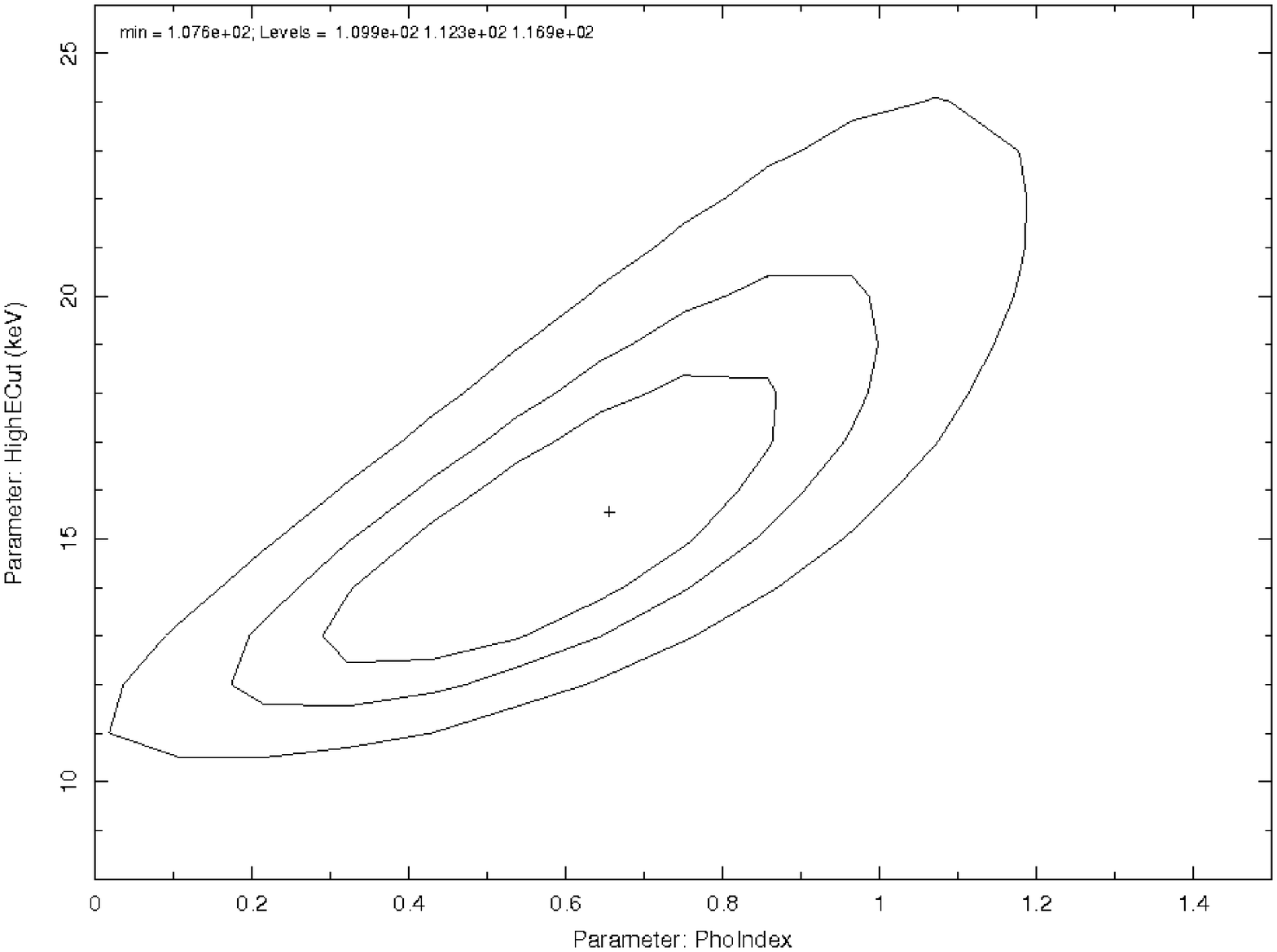}
\caption{Contour plots of a multi-parameter analysis of the cut-off power law model were produced with XSPEC.  The \textit{left} hand plot shows the correlation between N$_H$ and high energy cut-off; The \textit{right} hand plot shows the correlation between photon index and high energy cut-off.  Contours are shown at the 1, 2 and 3 $\sigma$ levels.}
\label{fig:specerr}
\end{figure*}

The 1--9 keV Swift/XRT spectrum and the 22--100 keV IBIS/ISGRI spectrum were simultaneously fit using the \emph{XSPEC} software, version 12.3. A systematic error of 2\% was included for the IBIS data and a cross-calibration constant was introduced into the model to account for any differences between the two instruments as well as to allow for source intensity variations considering the non-contemporaneous nature of the two spectra. 

An initial fit to the data was attempted using a simple absorbed power law, however the parameters of the fit were badly constrained and the reduced chi-square of the fit was very poor ($\chi^{2}_{\nu}$~=~1.8). Consequently more complex models were used.  The best fit ($\chi^{2}_{\nu}$~=~1.08) was achieved using an absorbed power law model with a high energy cut-off ({\it wabs*cutoffpl}) with a photon index of 0.6~$\pm$~0.3, a cut-off energy of 17$^{+5}_{-3}$ keV and N$_H$ of 5.4$^{+1.3}_{-1}$~$\times$~10$^{22}$~cm$^{-2}$ (see Figure~\ref{fig:spec}).  A cross calibration constant of $\sim$0.18 is required in the best fit model.  An absorbed thermal bremsstrahlung model was
also fit but was a worse fit to the data especially at low energies ($\chi^{2}_{\nu}$~=~1.21). The residuals in the spectral fit around 6.5 keV indicated the possible presence of an Fe line or edge.  Attempts to add either of these components to the spectral model do not significantly improve the fit and have unconstrained errors.  The results of the fits to the data are shown in Table~\ref{tab:spec}.  

\begin{table*}
\centering
\caption{The results of the spectral fits to the XRT and ISGRI IGR J16493$-$4348 data in the 1--100 keV band.  All three models included absorption components.  All errors are quoted at the 90\% level for a single parameter of interest.}
\begin{tabular}{|l|lll}
\hline
Parameter & Power law &	Cut-off Power law & Thermal Bremsstrahlung \\
\hline
Calib. constant & 0.43$^{+0.21}_{-0.14}$	&	0.18$^{+0.08}_{-0.05}$ & 0.32$^{+0.1}_{-0.07}$ \\
N$_H$ & 11$\pm$1~$\times$~10$^{22}$cm	& 5.4$^{+1.3}_{-1.0}$~$\times$~10$^{22}$~cm$^{-2}$ & 7.9$^{+0.7}_{-0.6}$~$\times$~10$^{22}$~cm$^{-2}$\\
Photon index	&  2.1$\pm$0.2	&0.6$\pm$0.3 & - \\
kT		& -	&	-	&	30$^{+8}_{-6}$ keV\\
E$_{cut}$ & -	& 17$^{+5}_{-3}$ keV & - \\
Normalisation	&	0.04$^{+0.02}_{-0.01}$ & 4.0$^{+2}_{-1}$~$\times$~10$^{-3}$ & 0.019$\pm$0.001 \\
$\chi$$^{2}$~/~dof	&	165.04/92	&	98.23/91	& 111.10/92\\
\end{tabular}
\label{tab:spec}
\end{table*}  

The main parameters of the cut-off power law model were investigated for correlations, specifically the N$_H$, cut-off energy and photon index.  Figure~\ref{fig:specerr} indicates the correlations between N$_H$ and photon index with the high energy cut-off.  In both cases it is clear that the parameters are bounded and constrained.  Additionally it is clear that within 3$\sigma$ the high energy cut-off can vary between $\sim$10 keV and $\sim$24 keV, i.e. the energy gap between the Swift/XRT and the INTEGRAL/IBIS data.  This uncertainty in the cut-off obviously impacts upon the index of the power law slope, requiring a much harder spectrum for a low cut-off energy than a higher one.  The absorbing column is consequently also correlated but within 3$\sigma$ lies in the range 3.8~$<$~N$_H$~$<$~8.1~$\times$~10$^{22}$~cm$^{-2}$.

The cross-calibration constant of the model is rather low indicating that the source is variable between the non-contemporaneous X-ray and soft $\gamma$-ray observations.  An additional explanation would be that the Swift-ISGRI association is spurious; this appears highly unlikely as IGR J16493$-$4348 is seen as a weak persistent source in the $\gamma$-ray band and both Chandra and Swift/XRT independently find an X-ray source at the same location within the INTEGRAL error circle.  No other X-ray source is a candidate as the X-ray counterpart.
  
The 1--100 keV flux for the source assuming the cut-off power law model is $\sim$1.1~$\times$~$10^{-10}$erg~cm$^{-2}$~s$^{-1}$; the fluxes in the 2--10 keV, 10--20 keV and 20--40 keV band are $\sim$5.3~$\times$~$10^{-11}$, $\sim$2.7~$\times$~$10^{-11}$ and $\sim$1.7~$\times$~$10^{-11}$~erg~cm$^{-2}$~s$^{-1}$ respectively.

\subsection{IR archive data}\label{sec:ir}
\citet{ATel654} reported that there was no identifiable optical counterpart to IGR J16493$-$4348 in the Digital Sky Survey Maps.  An examination of the USNO-B1.0 catalogue finds no optical counterparts within 10$\arcsec$ of the Chandra position.  Additionally, \citet{ATel654} report a single 2MASS source, 2MASS J16492695$-$4349090, compatible with both the Chandra and Swift/XRT positions at distances of 0.39$\arcsec$ and 4$\arcsec$ respectively.  It is the only 2MASS source within 7$\arcsec$ of the Chandra position \citep{2006AJ....131.1163S}.  \citet{ATel654} observed the source in the Ks band and found a magnitude of 12$\pm$1, consistent with the 2MASS magnitude and indicating that there is no large scale variability of the source at this band.

A single source in the Spitzer GLIMPSE catalogue, SSTGLMC G341.3752+00.5829, is compatible with the high energy observations.  This source lies 0.04$\arcsec$ from the position of the 2MASS counterpart; the GLIMPSE point source accuracy is reported to be typically 0.3$\arcsec$.

\begin{figure}
\centering
\includegraphics[width=0.95\linewidth, clip]{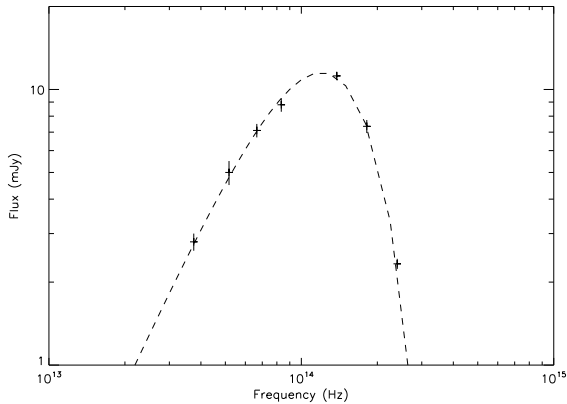}
\caption{The infra-red SED for IGR J16479$-$4348.  The fluxes are taken from the 2MASS and Spitzer GLIMPSE surveys (see Table~\ref{tab:ir}).  Fluxes have not been corrected for extinction in the line of sight or intrinsic to the source.  The dashed line represents the absorbed blackbody model fit discussed in $\S$~\ref{sec:ir}.}
\label{fig:sed}
\end{figure}

\begin{figure}
\centering
\includegraphics[width=0.95\linewidth, clip]{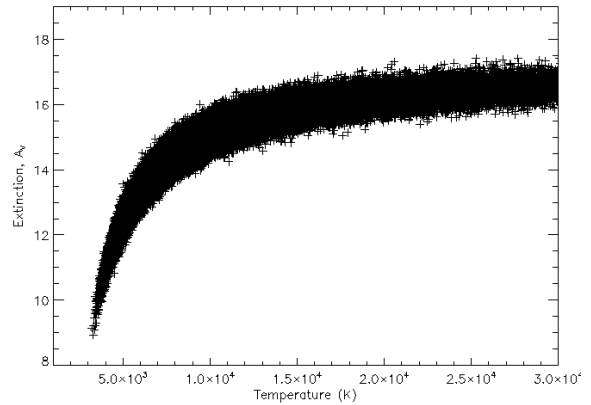}
\caption{A plot of the values of the Temperature and A$_V$ values fit to the Monte-Carlo simulated SEDs.  The points all occupy a fairly well defined parameter space.}
\label{fig:mc_res}
\end{figure}

The infra-red spectral energy distribution (SED) of the optical counterpart is shown in Figure~\ref{fig:sed}.  We fit the SED with a blackbody at temperature T with extinction A$_V$.  Assuming the source is a sphere of radius r and at a distance D from the Earth then the model flux is:
\begin{eqnarray}
f(v)	& = & \frac{2\times 10^{29} \pi h}{c^2}\frac{r^2}{D^2}\frac{\nu^3}{\exp{(h\nu /kT)} - 1}10^{-0.4A_{\nu}} \ \ \ \ mJy
\label{eqn:model}
\end{eqnarray}
where A$_{\nu}$ is the absorption at frequency $\nu$, given A$_V$, based upon the formulae of \citet{1989ApJ...345..245C}.  The model is successfully fit to the data with a reduced chi-square of $\chi$$^2_{\nu}$ = 0.45; the fit is shown in Figure~\ref{fig:sed} with parameters r$/$D =6.9~$\times$~10$^{11}$, T = 9166 K and A$_V$ = 14.8.  However, as noted by \citet{2004ApJ...616..469F} when performing a similar analysis, because of very strong parameter degeneracy this is not the only choice of viable model parameters.  To explore this degeneracy we adopt a Monte-Carlo approach whereby a simulated SED is generated by randomly varying the original SED within its statistical errors and refitting the absorbed black-body model. A good fit with physically meaningful parameters was achieved for $\sim$200,000 simulations and the results for the extinction and temperature parameters are plotted in Figure~\ref{fig:mc_res} and clearly indicate the degeneracy of the two parameters.

As observed by \citet{2004ApJ...616..469F}, in the analysis of IGR J16318$-$4848, two regions can be loosely defined corresponding to two different physical systems:
\begin{enumerate}
	\item A low temperature of 3000--10,000 K with a wide possible range of extinctions, 9~$<$~A$_V$~$<$~15.  The r$/$D ratio lies within the range (0.7--1.2)$\times$10$^{-10}$.  This is compatible with a main sequence dwarf star at a distance of 200--300 pc, or a cool red giant which if the star had a radius of $\sim$10 R$_{\sun}$ would be 2--3 kpc away.
	\item A high temperature of $>$11,000 K with an extinction constrained within 14.5~$<$~A$_V$~$<$~17.5.  The r$/$D ratio lies within the range (3--6)$\times$10$^{-11}$.  This is compatible with an early type stellar photosphere which assuming a stellar radius of 20--30 R$_{\sun}$ would place it 7.5--22 kpc away.
\end{enumerate}
The extinction can be intrinsic to the source, along the line of sight or a combination of both.

\begin{table}
\centering
\caption{The IR magnitudes of the 2MASS source, 2MASS J16492695$-$4349090, and the Spitzer source, SSTGLMC G341.3752+00.5829, which are discussed in the text as the counterpart to IGR J16493$-$4349}
\begin{tabular}{|l|l}
\hline
Waveband & Magnitude \\
\hline
J & 14.59 $\pm$ 0.05 \\
H & 12.86 $\pm$ 0.06 \\
Ks & 11.94 $\pm$ 0.04 \\
3.6$\mu$m & 11.25$\pm$0.07 \\
4.5$\mu$m & 11.01$\pm$0.07 \\
5.8$\mu$m & 10.9$\pm$0.1 \\
8.0$\mu$m & 10.86$\pm$0.09
\end{tabular}
\label{tab:ir}
\end{table}  
  
\section{Discussion}
The original association of IGR J16493$-$4349 and PSR J1649$-$4349 was based upon a low significance detection of a $\gamma$-ray source by the INTEGRAL satellite in the vicinity of PSR J1649$-$4349 \citep{Bird04-cat1, ATel457}.  The accumulation of more observations of this field, by INTEGRAL, has yielded the more significant source detection reported with a better fitted position and a smaller error circle which place the radio pulsar outside of the 90\% error circle.  There is a $<$1\% probability that the INTEGRAL/IBIS source position is compatible with the pulsar.  The Swift/XRT position further refines the location of the source to within an error circle of 3.6$\arcsec$.  Both of these positions are compatible with the Chandra position reported by \citet{ATel654} and imply the association with PSR J1649$-$4349 as spurious.  Furthermore the luminosity of IGR J16493$-$4348 if it were at the 5.6 kpc distance of the pulsar would be approximately five orders of magnitude higher than would be predicted from magnetic dipole losses.

The 1--100 keV broadband spectrum is fit best by an absorbed power law model with a high energy cut-off.  The reported spectral model and fit parameters (see Table~\ref{tab:spec}) are typical of accreting neutron stars in high mass X-ray binaries, e.g. \citet{2004A&A...418..655N}.  The RXTE-PCA spectral measurement by \citet{ATel465} is consistent with an absorbing column of, N$_H$~$\sim$~$10^{23}$cm$^{-2}$ this is higher than measured by Swift-INTEGRAL and may represent variation in the intrinsic absorption however, the non-imaging nature of the PCA instrument means that the spectrum was likely contaminated by 1RXS J164913.6-435527 \citep{ATel654}.  The RXTE/PCA 20--40 keV flux is given as 2.1~$\times$~10$^{-11}$~erg~cm$^{-2}$~s$^{-1}$ \citep{ATel465} this is comparable with the IBIS flux measured in this band of $\sim$1.7~$\times$~$10^{-11}$erg~cm$^{-2}$~s$^{-1}$.  However, the flux in the RXTE/PCA 2--10 keV band is a factor of $\sim$5 lower than that observed by Swift/XRT, despite the possible contamination from 1RXS J164913.6-435527 being in the field.  This may indicate that the soft X-ray flux is significantly variable whilst the harder X-ray flux is less so, however it must be remembered that the RXTE, Swift and INTEGRAL observations are not contemporaneous.  The Chandra flux is quoted as 0.255 counts s$^{-1}$ \citep{ATel654}; using the spectral parameters quoted in this paper and the HEASARC {\sc webpimms} tool this equates to a flux of $\sim$6.7~$\times$~$10^{-11}$erg~cm$^{-2}$~s$^{-1}$ in the 2--10 keV band.  This is comparable to the Swift/XRT flux although slightly brighter and provides further evidence that the source is variable in this energy band.

Estimates of the weighted average neutral hydrogen density in a cone of radius 1$^{\circ}$ along the line of sight of IGR J16493$-$4348 are N$_H$~$\sim$~1.4~$\times$~$10^{22}$cm$^{-2}$ \citep{2005A&A...440..775K} and N$_H$~$\sim$~1.8~$\times$~$10^{22}$cm$^{-2}$ \citep{1990ARA&A..28..215D}.  A range of (1.1--2.1)~$\times$~$10^{22}$cm$^{-2}$ is reported in the vicinity.  Assuming the maximum observed line of sight measurement and the 3$\sigma$ lower limit on the X-ray measured N$_H$ implies that the obscuration of the system is a minimum of 1.8 times that of the line of sight. This implies some level of absorption is intrinsic to the system.  Modelling the IR SED with an absorbed blackbody implies a range of extinctions of 9~$<$A$_V$~$<$17.5.  From \citet{1995A&A...293..889P} this equates to an N$_H$ of (1.6--3.1)~$\times$~10$^{22}$cm$^{-2}$; this is comparable to what is expected for the line of sight.  However, due to the degeneracy between the model parameters of SED model, the lower values of N$_H$ correspond are for cooler stars, which if they are on the main sequence, will be very close to us and consequently will effectively have no line of sight contribution; the higher N$_H$ values are slightly larger than the line of sight but would correspond to early type stars which would have a heavy stellar wind component which would increase their line of sight extinction.  The IR SED cannot explain the larger X-ray measured abosorbing column which would imply that any intrinsic obscuring material is concentrated around the compact object.

INTEGRAL has identifed a class of highly obscured HMXB sources with N$_H$~$>$~$10^{23}$cm$^{-2}$ \citep{2004ESASP.552..417W, 2005A&A...443..485D, 2006A&A...453..133W, 2007A&A...467..585B}.  While IGR J16493$-$4348 does not have an intrinsic absorption quite this high it does lie at the high end of the range of measured N$_H$ for traditional HMXB systems.  HMXB systems accreting via the stellar wind of a companion typically have luminosities of 10$^{35}$--10$^{36}$ erg s$^{-1}$ \citep{2004RMxAC..21..128K}. If the source is a HMXB then a likely location for the source would be within the Norma arm tangent, as this arm is tangential to the line of sight and which would give a source luminosity of $\sim$10$^{36}$ erg~s$^{-1}$. However, even if placed at 23 kpc away, on the far side of the galaxy, the source luminosity would not exceed $10^{37}$erg~s$^{-1}$ in the 1--100 keV band.  Furthermore studies of the luminosities of X-ray binaries broken down by compact object type have indicated that neutron star systems predominantly have luminosities of $\sim$$10^{36}$erg~s$^{-1}$ whilst black hole systems predominantly have luminosities of $\sim$$10^{37}$erg~s$^{-1}$ \citep{1996ApJ...473..963B, 2005A&A...443..485D}.  The range of reasonable luminosities for IGR J16439$-$4348 would therefore suggest it is more likely a neutron star system. 

Timing analysis of the INTEGRAL/IBIS data also shows no indication of a coherent timing signal on timescales of the order of 10--1000 s or of days to months.  The lack of any detected X-ray pulsations or any indication of orbital signature makes it difficult to say with any certainty that this is a high mass X-ray binary system.  However, there is evidence in the light curves that the source varies both in the X-ray and $\gamma$-ray bands.  This is further suggested by the low cross-calibration constant required in fitting the Swift-INTEGRAL spectra.

Despite the close proximity of the source to the plane of the Galaxy, the source lies only $\sim$20$^{\circ}$ away from the Galactic Centre and could conceivably be a low mass X-ray binary which is located towards the edge of the Galactic Bulge. Typical absorbing columns for LMXB sources are N$_H$~$<$~$10^{22}$cm$^{-2}$ \citep{1991ApJS...76.1127V, 2007A&A...467..585B}. Hence the identification of a column density higher than that associated with the galactic line of sight would make this a very unusual LMXB system.  Additionally, there is no indication in the 2.8 Ms of IBIS data of the source ever having undergone a type I X-ray burst which is characteristic of LMXB systems, however this is by no means conclusive.
 
\section{Conclusions}

We conclude that IGR J16493$-$4348 is not associated with the free radio pulsar, PSR J1649$-$4349 due to the position offset of the high energy emission and that the source is vastly overluminous in the 1--100 keV band to be powered purely by the spin-down of the pulsar.

The spectrum of IGR J16493$-$4348 is best fit by an absorbed power law with a high energy cut-off.  The spectral characteristics of the source are typical of an accreting neutron star.  The bright IR magnitudes of the IR counterpart together with the absence of an optical counterpart is indicative of an obscured source and are modeled well by an absorbed blackbody.  The location of the source in the region of the Norma arm spiral tangent is a tentative indicator that this source is likely to be a neutron star HMXB. The proximity of the source, however, to the Galactic Bulge means that a low mass X-ray binary nature of this source cannot be ruled out. 

The true nature of this source will only be definitively confirmed through further observations and study.  Specifically, IR spectroscopic follow-up may identify the nature of the donor star and any detection of X-ray pulsations would confirm the nature of the compact object.  However, based upon the location of the source, the possible intrinsic hydrogen obscuration and the X-ray spectra we suggest the object is likely to be an obscured HMXB accreting through a stellar wind.

\section*{Acknowledgments}

Based on observations with INTEGRAL, an ESA project with instruments
and science data centre funded by ESA member states (especially the PI
countries: Denmark, France, Germany, Italy, Switzerland, Spain), Czech
Republic and Poland, and with the participation of Russia and the USA.

This publication makes use of data products from the Two Micron All Sky Survey, which is a joint project of the University of Massachusetts and the Infrared Processing and Analysis Center/California Institute of Technology, funded by the National Aeronautics and Space Administration and the National Science Foundation.

We would like to thank the assistance of Nicola Masetti in the analysis of the IR data and also the anonymous referee for their constructive comments.

We acknowledge the following funding: in Italy, Italian Space Agency financial and programmatic support via contracts I/R/008/07/0 and I/023/05/0; in the UK, funding via PPARC grant PP/C000714/1.

\label{lastpage}

\end{document}